\documentclass{jaa}
\usepackage{natbib}
\usepackage{multirow}

\usepackage{graphicx}

\begin{document}\sloppy

\title{Detection of X-ray polarization in the high synchrotron peaked blazar
1ES 1959+650}


\author{Athira M. Bharathan\textsuperscript{1,*}, C. S. Stalin\textsuperscript{2}, Rwitika Chatterjee\textsuperscript{3}, S. Sahayanathan\textsuperscript{4,5}, Indrani Pal\textsuperscript{2}, Blesson Mathew\textsuperscript{1} and Vivek K. Agrawal\textsuperscript{3}}
\affilOne{\textsuperscript{1}Department of Physics and Electronics, CHRIST (Deemed to be University), Bangalore, India.\\}
\affilTwo{\textsuperscript{2}Indian Institute of Astrophysics, Block II, Koramangala, Bangalore 560 034 India.\\}
\affilThree{\textsuperscript{3}Space Astronomy Group, ISITE Campus, U. R. Rao Satellite Centre, Bangalore 560 037, India.\\}
\affilFour{\textsuperscript{4}Astrophysical Sciences Division, Bhabha Atomic Research Centre, Mumbai-400085, India.\\}
\affilFive{\textsuperscript{5}Homi Bhabha National Institute, Mumbai-400094, India.\\}

\twocolumn[{

\maketitle

\corres{athirabharathan1997@gmail.com}


\begin{abstract}
 We report the measurement of X-ray polarization in the 
high synchrotron peaked blazar 1ES 1959+650. Of the four epochs
of observations from the {\it Imaging X-ray Polarimetry Explorer}, we detected polarization in the 2$-$8 keV band 
on two epochs. From model-independent analysis of the observations on 28 October 2022, in the 2$-$8 keV band, we found the degree of polarization of $\Pi_X$ = 9.0 $\pm$ 1.6\% and an electric vector position angle of $\psi_X$ = 53 $\pm$ 5 deg. Similarly, from the observations on 14 August 2023, we found $\Pi_X$ and $\psi_X$ values as 12.5 $\pm$ 0.7\% and 20 $\pm$ 2 deg, respectively. These values are also in agreement with the values obtained from spectro-polarimetric analysis of the I, Q, and U spectra. The measured X-ray polarization is larger than the reported values in the optical, that ranges between 2.5$-$9\% , when observed during 2008 to 2018. Broadband spectral energy distribution constructed for the two epochs are well described by the one zone leptonic emission model with the bulk Lorentz factor ($\Gamma$) of the jet larger on 14 August 2023 compared to 28 October 2022. Our results favour shock acceleration of the particles in the jet, with the difference in $\Pi_X$ between the two epochs being influenced by change in the $\Gamma$ of the jet.
\end{abstract}


}]


\doinum{12.3456/s78910-011-012-3}
\artcitid{\#\#\#\#}
\volnum{000}
\year{0000}
\pgrange{1--}
\setcounter{page}{1}
\lp{1}

\section{Introduction}
Blazars are a peculiar category of active galactic nuclei (AGN)
that are powered by the accretion of matter onto super massive black hole
(SMBHs; 10$^6$ $-$ 10$^{10}$ M$_{\odot}$) residing at the center of
galaxies \citep{1969Natur.223..690L,1973A&A....24..337S,1984ARA&A..22..471R,
1993ARA&A..31..473A,1995PASP..107..803U}.
These blazars with their relativistic
jets oriented close to the line of sight to the observer emit
over the entire electromagnetic spectrum from the low energy radio to the
high energy $\gamma$-rays \citep{2019NewAR..8701541H}. 
The emission from these sources is primarily dominated by boosted non-thermal emission from particles in their relativistic jets. However, there is also a sub-dominant thermal emission component from the accretion disk and emission lines from their broad line region (BLR; \citealt{2020Galax...8...72C}).
The broadband spectral energy distribution (SED) of these sources shows a double hump structure. The low energy hump that peaks between the UV and X-ray frequencies is attributed to the synchrotron emission from relativistic electrons that gyrate along the magnetic field in the relativistic jets. The high energy hump peaks in the GeV to TeV energy
range and the physical process that gives rise to the high energy
hump is debated. Both leptonic and hadronic models are proposed to
explain the high energy emission process in blazars, however, we still
do not understand the origin of high energy emission in them.

In the leptonic model of blazar jet emission, the high energy emission is attributed to inverse Compton scattering of low energy photons by relativistic jet electrons. 
The seed photons can be the same synchrotron photons produced by the electrons via the synchrotron mechanism a process called synchrotron self Compton scattering (SSC; \citealt{1992ApJ...397L...5M,1997A&A...320...19M}). 
Alternatively, the seed photons can originate from a region external to the jet such as the accretion disk, the BLR and
the torus via a process called the external Compton scattering (EC; \citealt{1992A&A...256L..27D,
1994ApJ...421..153S}). Alternatively,
in the hadronic scenario, the high energy emission can be due to
synchrotron radiation by relativistic protons and photon-pion production
\citep{1992A&A...253L..21M,1993A&A...269...67M,2013ApJ...768...54B}. 
Based on
the peak frequency ($\nu_p$) of the synchrotron component of the broadband
SED, blazars are further classified into low synchrotron peaked (LSP; $\nu_p < 10^{15}$ Hz),
intermediate synchrotron peaked (ISP; $10^{14} < \nu_p < 10^{15}$ Hz) and high 
synchrotron peaked (HSP; $\nu_p > 10^{15}$ Hz) blazars.

Leptonic, hadronic, and/or lepto-hadronic models are able to provide a good fit to 
the broadband SED of blazars \citep{2013ApJ...768...54B,2018ApJ...863...98P}. 
A key observable to differentiate between these two models for explaining the broadband SED is through flux variability observations across different wavelengths.
Ideally, correlated optical and $\gamma$-ray flux variations will favor a leptonic scenario. This is due to the fact that a $\gamma$-ray flare without a counterpart at lower energies in the optical/IR 
cannot be produced in the one zone leptonic scenario. On the other hand, the correlations between optical and $\gamma$-ray flux variations are observed to be complex.\citep{2019MNRAS.486.1781R,2020MNRAS.498.5128R,2021MNRAS.504.1772R}. 
Another
observable that can constrain these two models is X-ray polarization 
\citep{2013ApJ...774...18Z,2022ApJ...931...59P}. Also, X-ray polarization
can provide strong signatures of the particle acceleration processes \citep{2022Galax..10..105S}.

Optical polarization is a characteristic feature of blazars, and monitoring observations have revealed that polarization variations often correlate with flares at other wavelengths \citep{1966ApJ...146..964K,1980ARA&A..18..321A,2016ApJ...833...77I,2017ApJ...835..275R,2022MNRAS.517.3236R,2022MNRAS.510.1809P}.
Polarization observations can thus be 
the key to the emission processes in the central regions of blazars as well as provide the diagnosis of the
orientation of the magnetic field in the jet. Though blazars
have been studied extensively for polarization in the optical and radio regimes, there was a dearth of X-ray polarization studies till the launch of the {\it Imaging X-ray Polarimetry 
Explorer (IXPE;}\citealt{2022JATIS...8b6002W}) in December 2021. 
Since its launch, {\it IXPE} has observed both Seyfert type AGN as well
as blazars. As of today, X-ray polarization characteristic is
known for about a dozen blazars, namely, Mrk 421 \citep{2023arXiv231006097K,2023NatAs...7.1245D}, 
1ES 0229+200 \citep{2023arXiv231001635E}, 
PG 1553+113 \citep{2023ApJ...953L..28M}, BL Lacertae \citep{2023ApJ...948L..25P,2023ApJ...942L..10M}, 
 Mrk 501 \citep{2022Natur.611..677L,2024ApJ...970L..22H,2024arXiv240711128C}, 3C 273, 3C 279, 3C 454.3 and S5 0716+714 \citep{2023arXiv231011510M}.

In the context of this work it is necessary to understand previous studies pertaining to the object of interest, 1ES 1959+650, which is a BL Lac object at a redshift of z = 0.048 \citep{1996ApJS..104..251P}. This source is a HSP blazar whose synchrotron peak in the broadband SED falls in the UV to X-ray region \citep{2004ApJ...601..151K,2016MNRAS.461L..26K}. 
It was first detected in the radio band using the \textit{Green Bank Telescope} \citep{1991ApJS...75.1011G}. In X-rays, it was initially identified in the Einstein Slew Survey \citep{1992ApJS...80..257E} and later observed with \textit{BeppoSAX}, \textit{Swift}, and \textit{XMM-Newton} \citep{2002A&A...383..410B,2008A&A...478..395M}. Additionally, it was detected in the GeV–TeV energy range \citep{1999ICRC....3..370N,2003ApJ...583L...9H}.
This source was found to show uncorrelated flux variations between
X-rays and TeV $\gamma$-rays \citep{2004ApJ...601..151K,2014ApJ...797...89A},
which is inconsistent with the simple one
zone SSC model. Alternatively, such uncorrelated flux variations can be
explained by multi-zone SSC mechanism \citep{2008ApJ...689...68G},
EC mechanism \citep{2004ApJ...601..151K} or hadronic processes
\citep{2005ApJ...621..176B}. 1ES 1959+650 was observed by {\it IXPE} during four epochs, between May 2022 and August 2023, and in this work we report the measurements of X-ray polarization in 1ES 1959+650.

Broadband SED modeling of blazars can provide important constraints
on the physical processes happening close to their central regions.
To look for any correlation between the physical parameters deduced
from SED modeling and the observed X-ray polarization, we also 
generated the broadband SEDs of the source using near-simultaneous
data acquired at optical, UV, X-ray and $\gamma$-ray energies. The 
SEDs thus generated were modeled using simple one-zone leptonic
emission model. The paper is organized as follows: In Section 2, we describe the observations and reduction of data, analysis and results are detailed in Section 3, the discussion in Section 4, followed by the conclusions in the final Section.

\section{Observations and data reduction}
\label{obs}
\subsection{X-ray polarization}
{\it IXPE} observed 1ES 1959+650 during four epochs between May 2022 and August 2023 with its three detector units (DU) for a net exposure ranging between $\sim$50 
and $\sim$300 ksec. The log of the {\it IXPE} observations is given in 
Table \ref{table-1}. 

\begin{table}[htb]
\tabularfont
\caption{The log of {\it IXPE} observations. The details are the observational ID (OBSID), Date of observation and the exposure time.}\label{table-1} 
\begin{tabular}{lcc}
\topline
OBSID& Date &Exposure Time (secs) \\\midline
01006201 & 03 May2022 & 53519 \\
01006001 & 09 June 2022 & 200432 \\
02004801 & 28 October 2022 & 194844 \\
02250801 & 14 August 2023 & 312477 \\
\hline
\end{tabular}
\end{table}

\begin{figure*}
\centering
\includegraphics[scale=0.7]{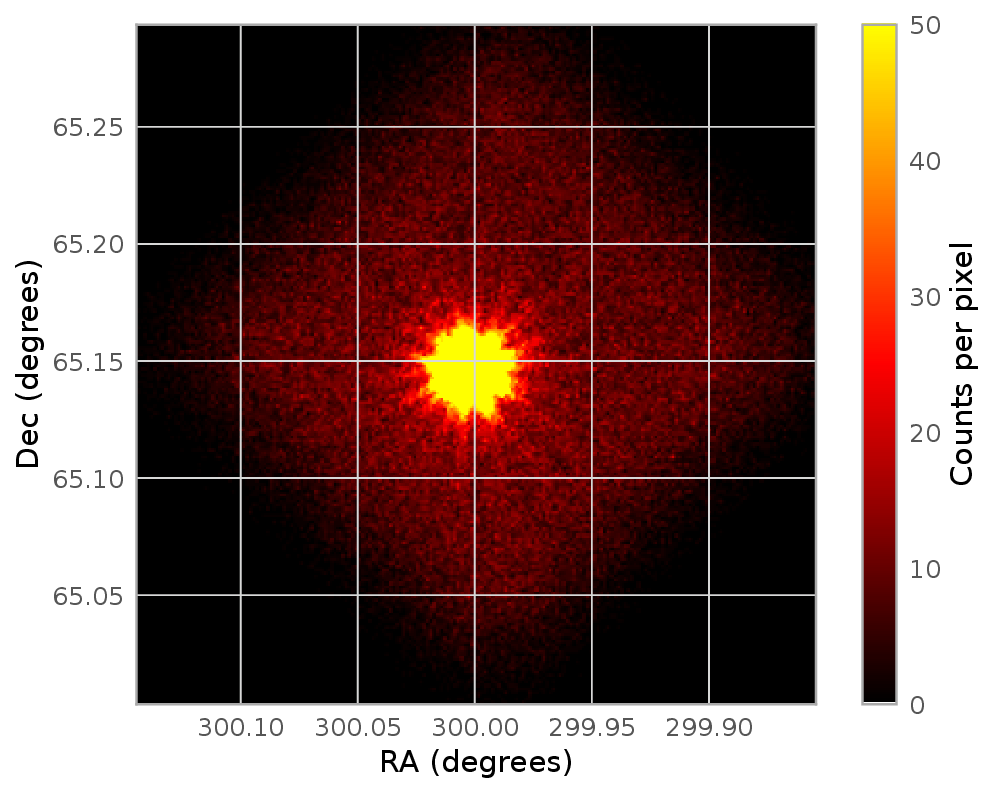}
\caption{ The generated count map in sky coordinates using the CMAP algorithm within the xpbin task.}
\label{figure-1}
\end{figure*}

We used the cleaned and calibrated level 2 data for the scientific analysis. We analyzed the publicly available level 2 data using {\tt IXPEOBSSIM} software 
v30.0.0 \citep{2022SoftX..1901194B}. We generated a count map (see Fig. \ref{figure-1}) in sky coordinates 
using the {\tt CMAP} algorithm within the {\tt xpbin} task. We adopted a 
circular region with a radius of $70''$ for the source extraction from the 
three DUs, and a source-free region with a radius of $100''$ for the 
background extraction for each DU. We then used the {\tt xpselect} task to 
generate the filtered source and background regions for the polarimetric analysis. 
For spectro-polarimetric analysis, we generated the I, Q and U source and 
background spectra using the {\tt PHA1}, {\tt PHA1Q} and {\tt PHA1U} algorithms 
using {\tt xpbin} task within {\tt IXPEOBSSIM} for the three DUs. 

\subsection{{\it Fermi}-LAT}
The $\gamma$-ray data was obtained from the Large Area Telescope (LAT;\citealt{2009ApJ...697.1071A}) onboard the
{\it Fermi Gamma Ray Space Telescope} (hereinafter {\it Fermi}) 
launched in June 2008. We downloaded the Pass8 data from
the archives and followed the standard LAT data analysis procedure\footnote{https://fermi.gsfc.nasa.gov/ssc/data/analysis/} using {\it fermipy} \citep{2017ICRC...35..824W}. We used the data in the energy range of 100 MeV to 
300 GeV and considered all the SOURCE class events (evclas=128 and
evtype=3) in a region of interest (ROI) of 10$^{\circ}$ centred on
the source position. To select data with good time interval we
used the filter "DATA\_QUAL $>$ 0 \&\& LAT\_CONFIG==1". As background
models we used the galactic diffuse emission model
(gll\_iem\_v07) and isotropic diffuse emission model
(iso\_P8R3\_SOURCE\_V3\_v1) and carried out a binned
likelihood analysis to generate $\gamma$-ray spectra. In this work, we used data from {\it {Fermi}} for two epochs, one acquired during the period 27 $-$ 28 October 2022 and the other acquired during the period 13 $-$ 15 August 2023.

\subsection{Swift-XRT}
For X-rays, we used the data acquired by the X-Ray Telescope (XRT; \citealt{2005SSRv..120..165B}) onboard 
the Neil Gehrels Swift Observatory,\footnote{https://heasarc.gsfc.nasa.gov/cgi-bin/W3Browse/swift.pl} simultaneous with \textit{IXPE} observations on 28 October 2022 (OBSID $=$ 00096560021) and14 August 2023 (OBISD $=$ 00097164002).
We used the task {\it xrtpipeline}\footnote{https://www.swift.ac.uk/analysis/xrt/xrtpipeline.php}
from HEASOFT\footnote{https://heasarc.gsfc.nasa.gov/docs/software/heasoft/} to process the 
data from XRT. We used the energy range from 0.5 to 10 keV. We chose a circular
region of radius 10 arcsec for the source and for the background we selected a circular
region of radius 40 arcsec away from the source. We used
{\it xrtmkarf} to generate the ancillary response file and
used {\it grppha} to group the spectra. We grouped the 
spectra to have 20 counts per energy bin and fitted
the spectra with an absorbed power law model. While
fitting, we fixed the Galactic hydrogen column density ($N_{H}$)
to the value of 0.101$\times$10$^{22}$$cm^{-2}$ \citep{2013MNRAS.431..394W}.

\subsection{Swift-UVOT}
For optical and UV we used the data acquired on 28 October 2022 and 14 August 2023 from the Ultra-Violet Optical Telescope (UVOT; \citealt{2005SSRv..120...95R}) onboard 
the Neil Gehrels Swift Observatory. We used
data in three optical filters V, B and U bands
and two UV filters W1 and W2 bands and taken
from the archives at HEASARC\footnote{https://heasarc.gsfc.nasa.gov/cgi-bin/W3Browse/w3browse.pl}. 
We used
the tool '\textit{uvotsource}' to get the instrumental magnitudes.
They were corrected for galactic extinction using a E(B$-$V) of 0.178 from \cite{2011ApJ...737..103S} and the extinction laws of 
\cite{1989ApJ...345..245C}. The extinction corrected magnitudes
were then converted to flux densities using
the conversion factors given in \cite{2011AIPC.1358..373B}.

\section{Analysis and Results}
\label{sec:ar}
\subsection{Polarimetry}

\begin{figure*}
\centering
\includegraphics[scale=0.3]{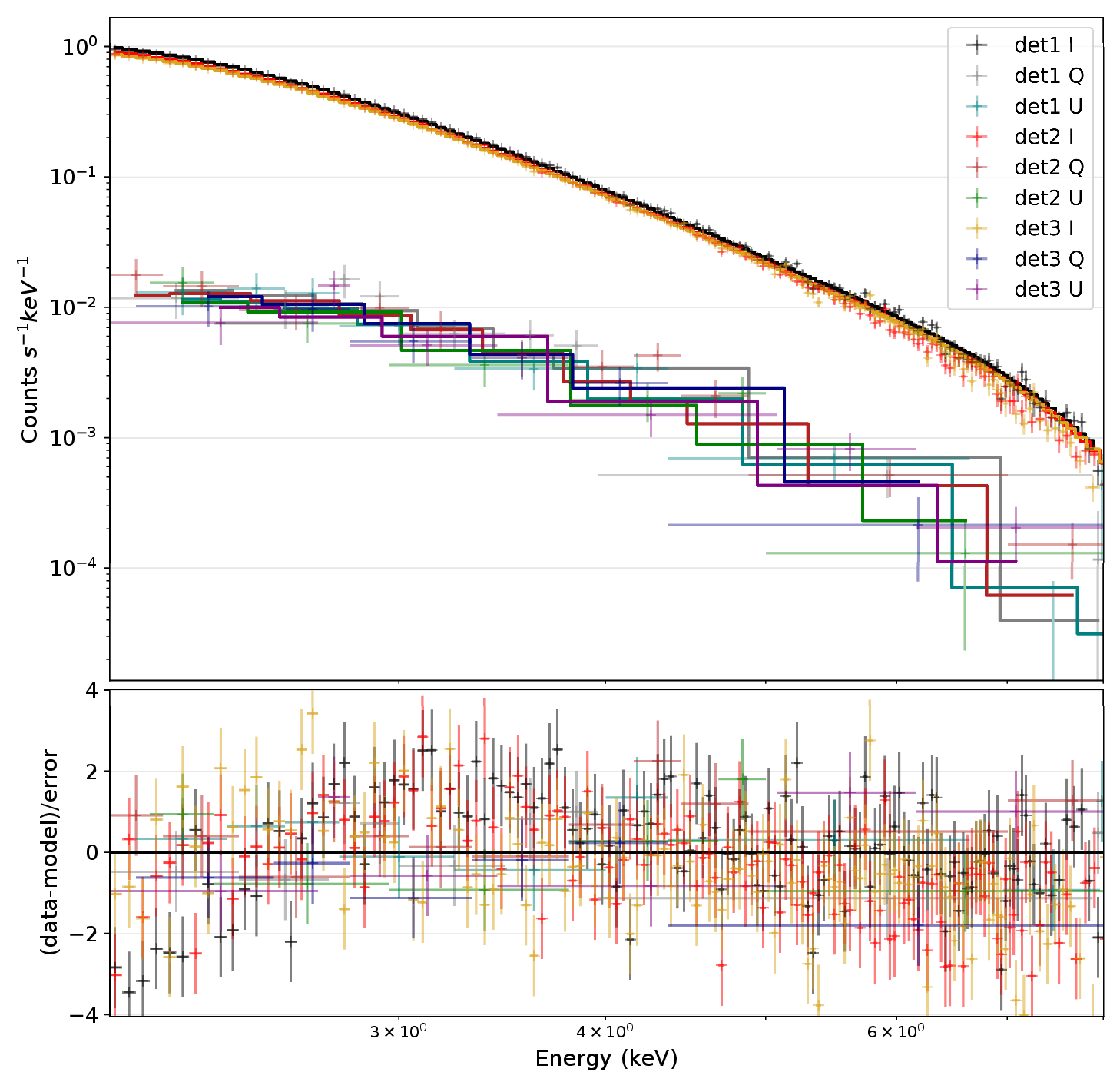}
\includegraphics[scale=0.36]{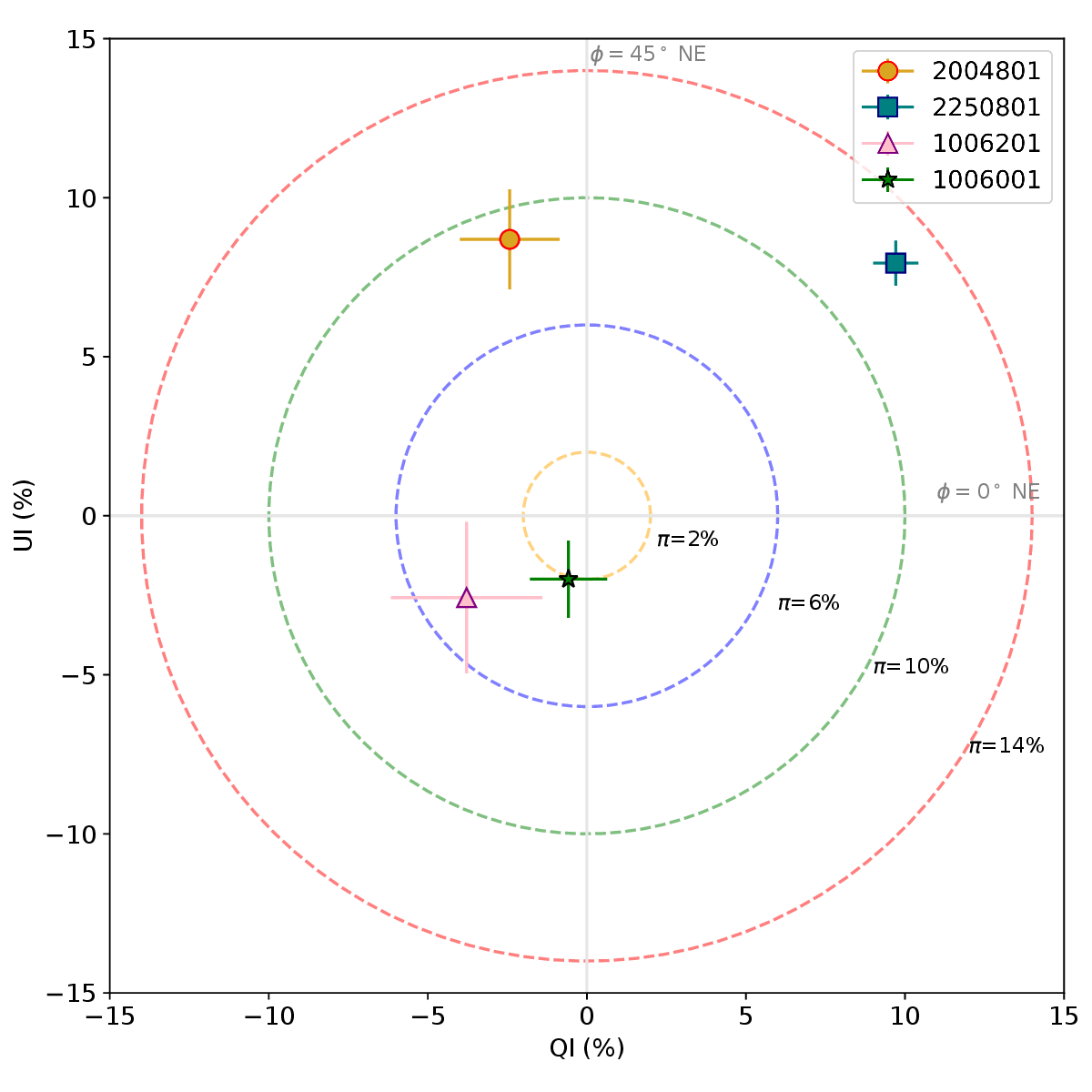}
 \caption{{\it IXPE} I, Q, U Stokes best-fit spectra with residuals in the 2$-$8 keV band for the observations of 14 August 2023 (left) and normalized U/I and Q/I Stokes parameter in the total 2$-$8 keV band of {\it IXPE} for 03 May 2022 (pink triangle), 09 June 2022 (green asterisk), 28 October 2022 (orange circle), and 14 August 2023 (blue square)(right). }
 \label{figure-2}
\end{figure*}

We analyzed the polarimetric signal from 1ES 1959+650 for all
the four OBSIDs using {\tt PCUBE} 
algorithm in the {\tt xpbin} task. We generated the three polarization 
cubes for the three DUs to extract information such as the Stokes parameters 
(I, Q, U), the minimum detectable polarization (MDP), the polarization degree 
($\Pi_{X}$), the polarization angle ($\Psi_{X}$) and their associated errors. We 
first generated the three polarization cubes corresponding to three DUs in the 
entire 2$-$8 keV energy band. The combined polarization parameters from the 
three DUs in the 2$-$3 keV, 3$-$5 keV, 5$-$8 keV and 2$-$8 keV bands are given in Tables \ref{table-2} and \ref{table-3}. Of the 
four OBSIDs we detected significant polarization on two epochs, namely,
28 October 2022 and 14 August 2023. In these two epochs, the measured
polarization is beyond the MDP values. 
The normalized U/I and Q/I Stokes 
parameters obtained from the combined cube for the observations of all the epochs are shown in Fig. \ref{figure-2}.

\begin{table*}[htb]
\tabularfont
\caption{The measured polarization parameters in different energy bands. The quoted errors in $\Pi_X$
are the 1 sigma errors.}\label{table-2} 
\begin{tabular}{cccccccccccc}
\topline
OBSID & Date & \multicolumn{4}{c}{$\Pi_X$ (\%)} & \multicolumn{4}{c}{MDP (\%)} \\\midline
& & 2$-$3 keV & 3$-$5 keV & 5$-$8 keV & 2$-$8 keV & 2$-$3 keV & 3$-$5 keV & 5$-$8 keV & 2$-$8 keV\\\midline
01006201 & 03 May 2022 & 3.7$\pm$3.1 & 9.5$\pm$2.8 & 12.3$\pm$6.8 & 4.6$\pm$2.4 & 9.3 & 8.6 & 20.5 & 7.2 \\
01006001 & 09 June 2022 & 2.1$\pm$1.5 & 2.9$\pm$1.5 & 3.5$\pm$3.8 & 2.1$\pm$1.2 & 4.7 & 4.5 & 11.4 & 3.7 \\
02004801 & 28 October 2022 & 7.8$\pm$1.9 & 11.7$\pm$1.9 & 8.4$\pm$5.2 & 9.0$\pm$1.6 & 6.0 & 6.1 & 15.7 & 4.7 \\
02250801 & 14 August 2023 & 11.8$\pm$0.1 & 12.6$\pm$0.9 & 15.9$\pm$2.6 & 12.5$\pm$0.7 & 2.7 & 2.8 & 7.9 & 2.1 \\
\hline
\end{tabular}
\end{table*}

\begin{table*}
\begin{center}
\tabularfont
\caption{The measured polarization parameters in different energy bands. The quoted errors in $\Pi_X$
are the 1 sigma errors.}\label{table-3} 
\begin{tabular}{cccccccc}
\topline
OBSID & Date & \multicolumn{4}{c}{$\psi_X$ (deg)}\\\midline
& & 2$-$3 keV & 3$-$5 keV & 5$-$8 keV & 2$-$8 keV \\\midline
01006201 & 03 May 2022 & $-$69$\pm$23 & $-$55$\pm$9 & 62$\pm$16 & $-$73$\pm$15 \\
01006001 & 09 June 2022 & $-$53$\pm$21 & $-$67$\pm$14 & $-$17$\pm$31 & $-$53$\pm$16 \\
02004801 & 28 October 2022 & 56$\pm$7 & 46$\pm$5 & 67$\pm$18 & 53$\pm$5 \\
02250801 & 14 August 2023 & 20$\pm$2 & 21$\pm$2 & 18$\pm$5 & 20$\pm$2 \\
\hline
\end{tabular}
\end{center}
\end{table*}

\begin{figure*}
\centering
\includegraphics[scale=0.35]{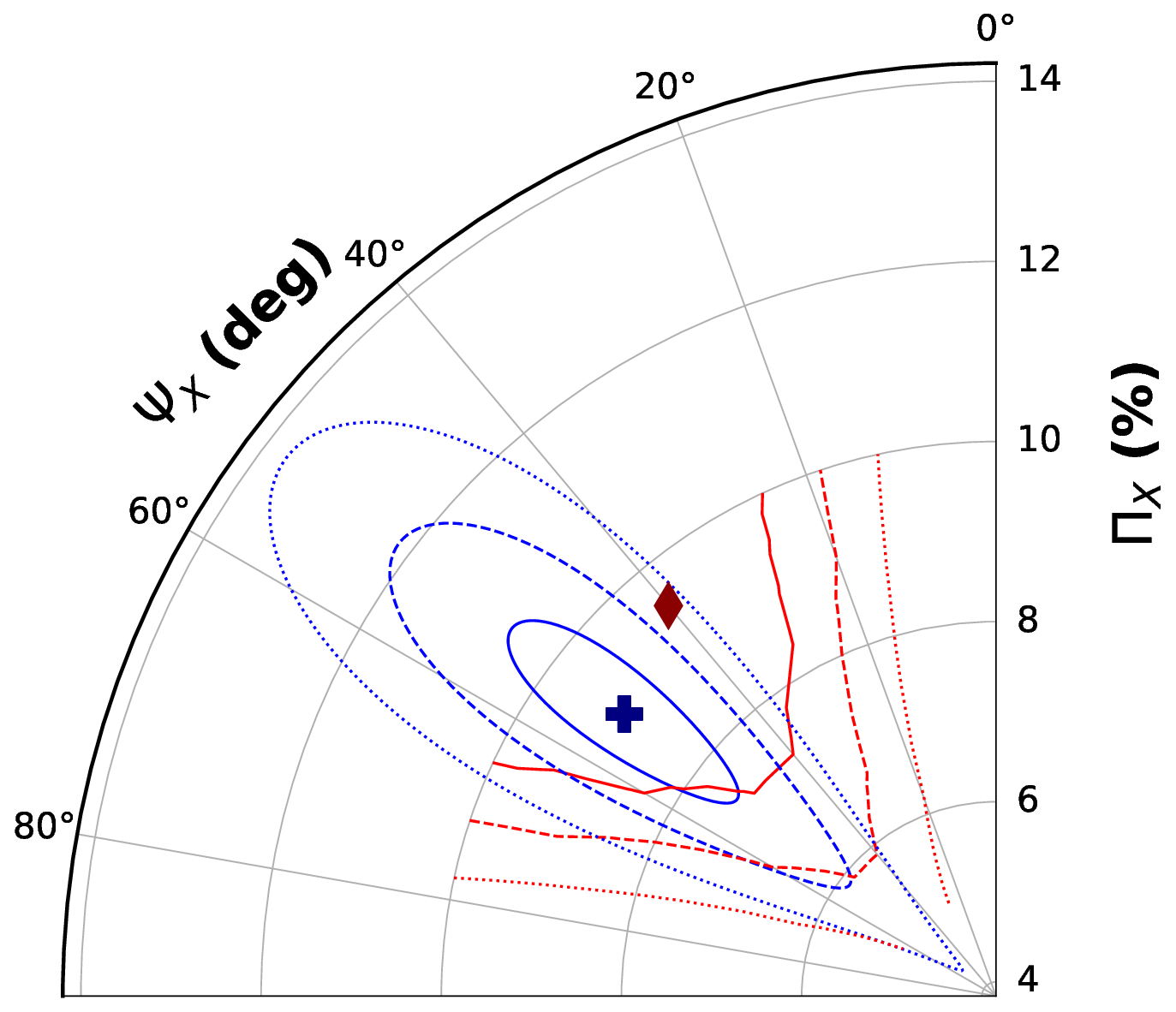}
\includegraphics[scale=0.35]{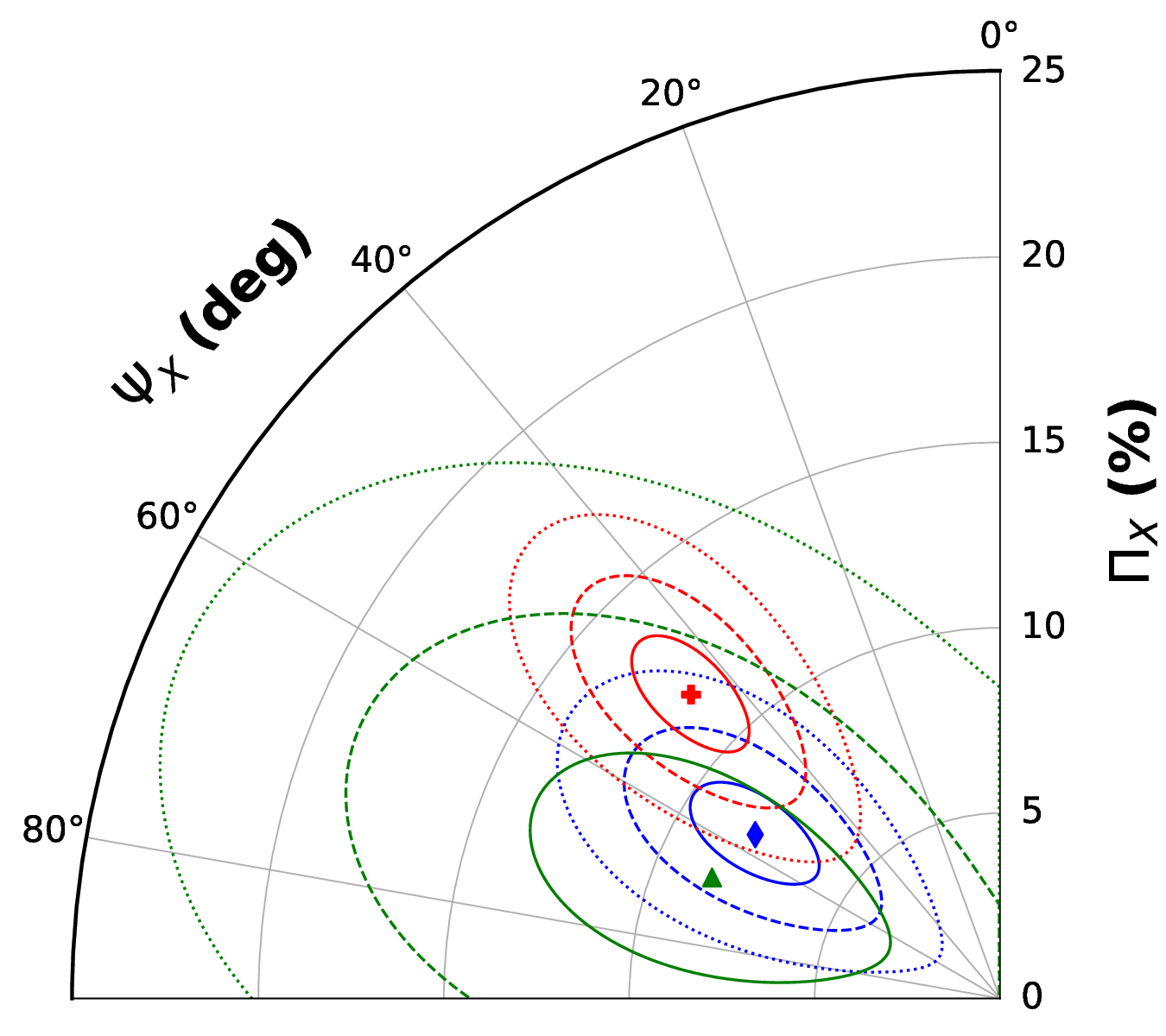}
\includegraphics[scale=0.35]{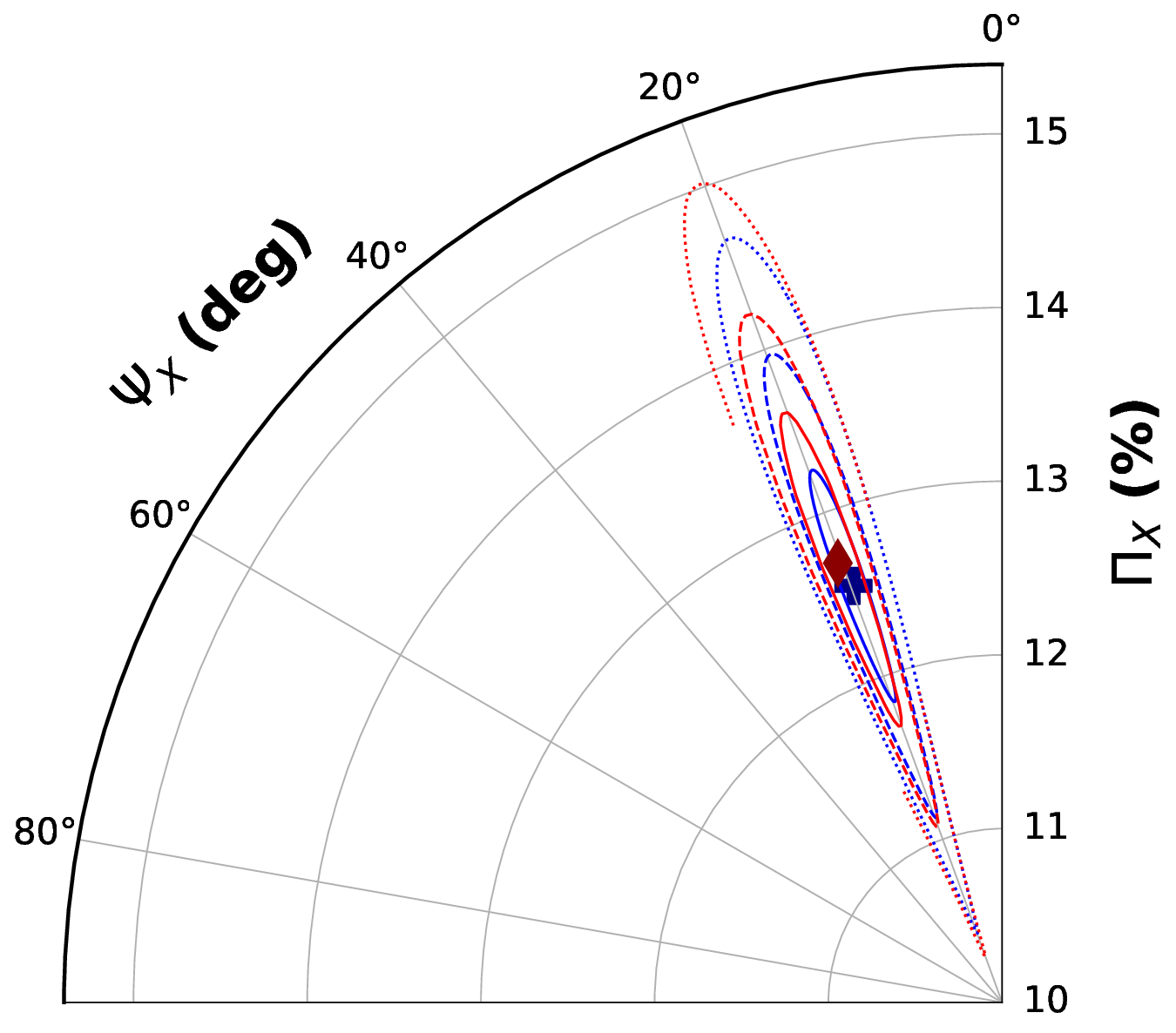}
\includegraphics[scale=0.35]{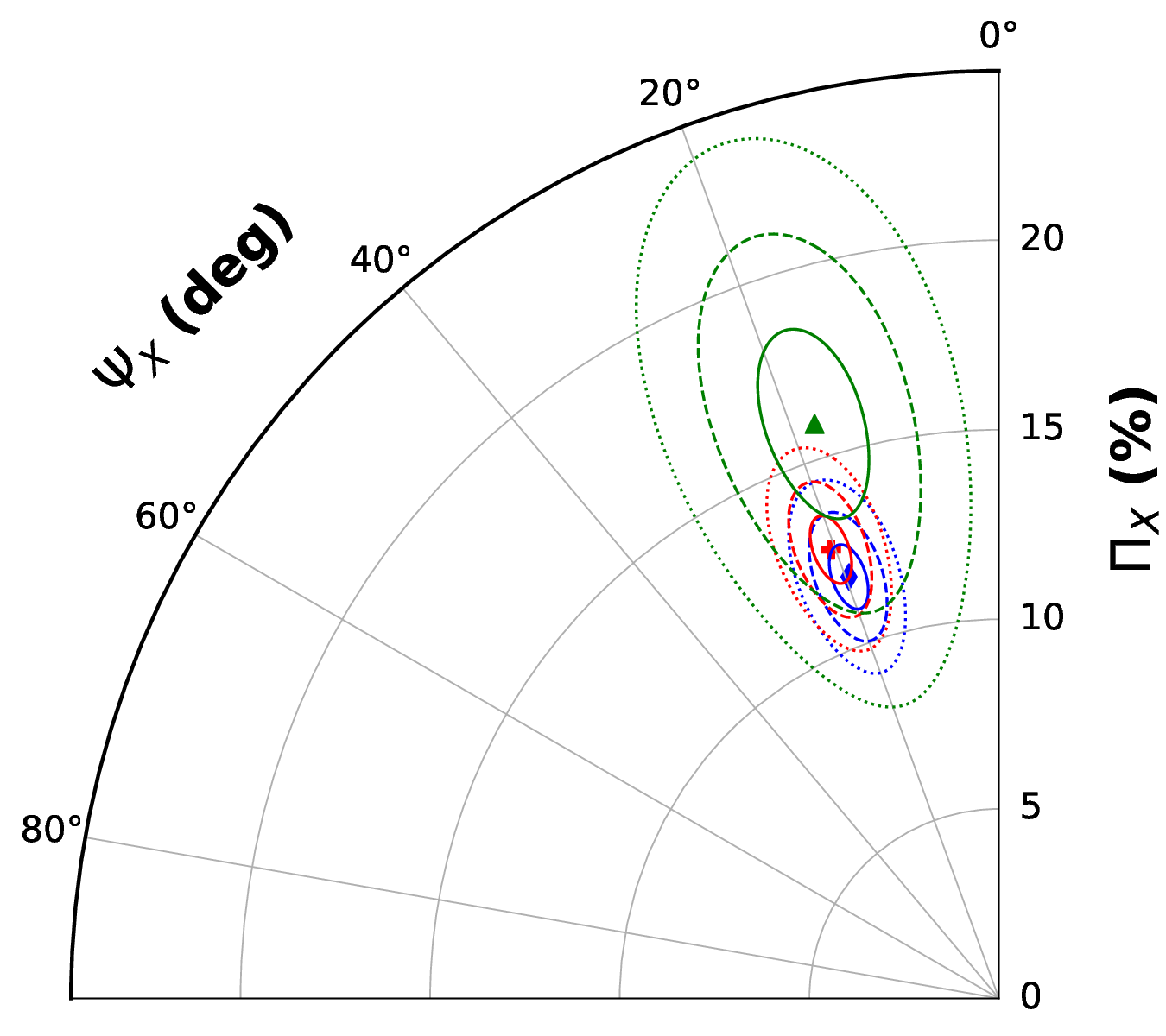}
\caption{The top panels display results from the observation conducted on October 28, 2022, while the bottom panels present data from the observation performed on August 14, 2023. The left panels show the measured polarization in the 2$-$8 keV band in the polarization degree ($\Pi_{X}$) and polarization angle ($\Psi_{X}$) plane, as determined by both PCUBE (blue) and XSPEC (red) analysis. The open contours for XSPEC analysis indicate that the specific orientation of the polarization vectors cannot be determined with confidence from the available data. The right panels display the measured polarization position across different energy bands. Diamond, plus, and triangle symbols denote the 2$-$3 keV, 3$-$5 keV, and 5$-$8 keV energy ranges, respectively. Additionally, the plot displays 68\%, 90\%, and 99\% confidence contours for  $\Pi_{X}$ and $\Psi_{X}$. }
\label{figure-1111}
\end{figure*}
To check for the energy dependence of the polarization parameters, we also derived 
the polarization parameters in three energy bins of 2$-$3 keV, 
3$-$5 keV and 5$-$8 keV using the {\tt PCUBE} algorithm. The derived parameters along with their 1 sigma errors
are given in Table \ref{table-2} and Table \ref{table-3}. The polarization values 
for the total energy range of 2$-$8 keV as well as for the
different energy ranges are given in Fig. \ref{figure-1111} for
both the epochs where significant polarization was detected. 
From Table \ref{table-2} and Table \ref{table-3} it is evident that we could not detect polarization in 1ES 1959+650 during the
observations carried out on 03 May 2022 and 09 June 2022. Polarization was notably identified in the subsequent two epochs of observations conducted on 28 October 2022 and 14 August 2023. On the 
observations of 28 October 2022, $\Pi_X$ and $\psi_X$ are
found to be similar (within errors) in the energy range of 2$-$3, 3$-5$ and the 2$-$8 keV. 
From the observations of 14 August 2023, we noticed that there is a tendency for an increase in $\Pi_{X}$ from the lower energy range to the
higher energy range, with $\psi_X$ remaining constant at $\sim$20 deg
between different energy bands. We note here, that Errando et al. (2024) based on spectro-polarimetric analysis of the data acquired on 03 May 2022, reported $\Pi_X$ = 8.0 ± 2.3\% and $\psi_X$ = 123 ± 8 degrees. Our non-detection of polarization on 03 May 2022 is likely due to our less sensitive model independent approach, first adopted on all the observations analysed in this work. However, our derived value of $\psi_X$ = $-$73 ± 15 deg (equivalent to 107 ± 15 deg) on 03 May 2022, is in agreement with the $\psi_X$ value reported by Errando et al. (2024) within errors.

\subsection{Spectro-polarimetry}
Among the four observational epochs, we could detect X-ray polarization on two epochs from model independent analysis. For those two epochs, we carried out the spectro-polarimetric analysis of the {\it IXPE} I, Q, U spectra in the 2$-$8 keV energy band. For the spectral fitting, we used an absorbed {\it powerlaw}, modified by a 
multiplicative constant polarization model {\it polconstant} in 
XSPEC \citep{1996ASPC..101...17A}. This model assumes constant polarization 
degree and a constant polarization angle over the specific energy band. In XPSEC 
the model takes the following form,
\begin{equation}
 constant \times TBabs \times (polconst \times pow)
\end{equation}

Here, {\it const} represents the inter-calibration constant for each detector. 
{\it TBabs} was used to model the Milky Way Galactic hydrogen column density, 
which was taken from \cite{2013MNRAS.431..394W}. We fixed $N_{H}$ during the fit. This model fits the I, Q, and U spectra (from the three detectors) 
well. The best fit I, Q and U
spectra along the residuals are given in Fig. \ref{figure-2}. All the model 
parameters for each of the detectors were tied during the fit. The errors were 
calculated at the 90$\%$ confidence ($\chi^2$ = 2.71 criterion). The best-fit 
parameters are given in Table \ref{table-4}.


\begin{table*}[htb]
\tabularfont
\caption{Comparison between the derived parameters via model independent and spectro-polarimetric analysis carried out in the 2$-$8 keV band for the two epochs where polarization is detected. Here, $\Pi_X$ is in percentage, $\psi_X$ is in degrees and $\Gamma$ is the photon index.The quoted errors in $\Pi_X$
are the 1 sigma errors.}\label{table-4} 
\begin{tabular}{cccccccccc}
\topline
OBSID & Date & \multicolumn{2}{c}{Model Independent} & \multicolumn{4}{c}{Spectro-polarimetry}\\\midline
& & $\Pi_X$ (2$-$8 keV) & $\psi_X$ (2$-$8 keV) & $\Pi_X$ (2$-$8 keV) & $\psi_X$ (2$-$8 keV) & $\Gamma$  & $\chi^2/dof$ \\\midline
02004801 & 28-10-2022 & 9.0$\pm$1.6 & 53$\pm$5 & 9.6$_{-0.1}^{+0.03}$ & 49$_{-32}^{+39}$ & 2.45$\pm$ 0.02 & 1.1 \\
02250801 & 14-08-2023 & 12.5$\pm$0.7 & 20$\pm$2 & 12.6$\pm$1.1 & 20$\pm$2 & 2.69$\pm$0.01  & 1.3 \\
\hline
\end{tabular}
\end{table*}

\subsection{Spectral Energy Distribution}
We modeled the SED of 1ES 1959+650 for the two epochs when X-ray
polarization was detected (namely 28 October 2022 and 14 August 2023) to check for any possible variation in the physical properties of the source which can result in X-ray polarization. The $\gamma$-ray data was obtained from the LAT onboard {\it Fermi} 
while the X-ray and optical/UV data were from \textit{Swift}-XRT and 
\textit{Swift}-UVOT telescopes (see Section 2). 
We modeled the observed SEDs using a simple leptonic model considering 
synchrotron and SSC emission mechanisms \citep{2018RAA....18...35S}.
In this model, a spherical region of size R is assumed to be the emission zone, which is filled with a distribution of non-thermal electrons, described by a broken power law of the form,
\begin{align} \label{eq:broken}
	N(\gamma)\,d\gamma = \left\{
\begin{array}{ll}
	K\,\gamma^{-p}\,d\gamma&\textrm{for}\quad \mbox {~$\gamma_{\rm min}<\gamma<\gamma_b$~} \\
	K\,\gamma_b^{q-p}\gamma^{-q}\,d\gamma&\textrm{for}\quad \mbox {~$\gamma_b<\gamma<\gamma_{\rm max}$~}
\end{array}
\right.
\end{align} 
Here, $p$ and $q$ are the low and high energy power-law indices, $\gamma$ is 
the electron Lorentz factor, and $\gamma_b$ is the break energy. The emission region is permeated with a tangled 
magnetic field $B$ and moves down the jet with a bulk Lorentz factor $\Gamma$. 

We found that 
the simple one zone leptonic model can reproduce the SED during both these states 
reasonably well. The optical/UV spectrum during the low flux state, 28 October 2023, shows significant deviation from a power law and better reproduced by thermal accretion disk with the inner disk temperature 0.4 eV. During this epoch, the X-ray spectrum is hard and significantly different from the $\gamma$-ray region. 
In our model the $\gamma$-ray spectral index is governed by the low energy index of the particle distribution while the X-ray spectral index is governed by the high energy index of the particle distribution.
Consistently the synchrotron spectrum predicted by the model is flat and declined sharply beyond X-ray energy. This decline is due to the maximum energy cutoff in the particle distribution. However there is no data to verify this sharp fall.
Alternatively, an exponential cutoff at $\gamma_{\rm max}$ can smoothen the steep spectral decline.

\begin{figure*}
\centering
\hbox{
      \includegraphics[scale=0.4,angle=270]{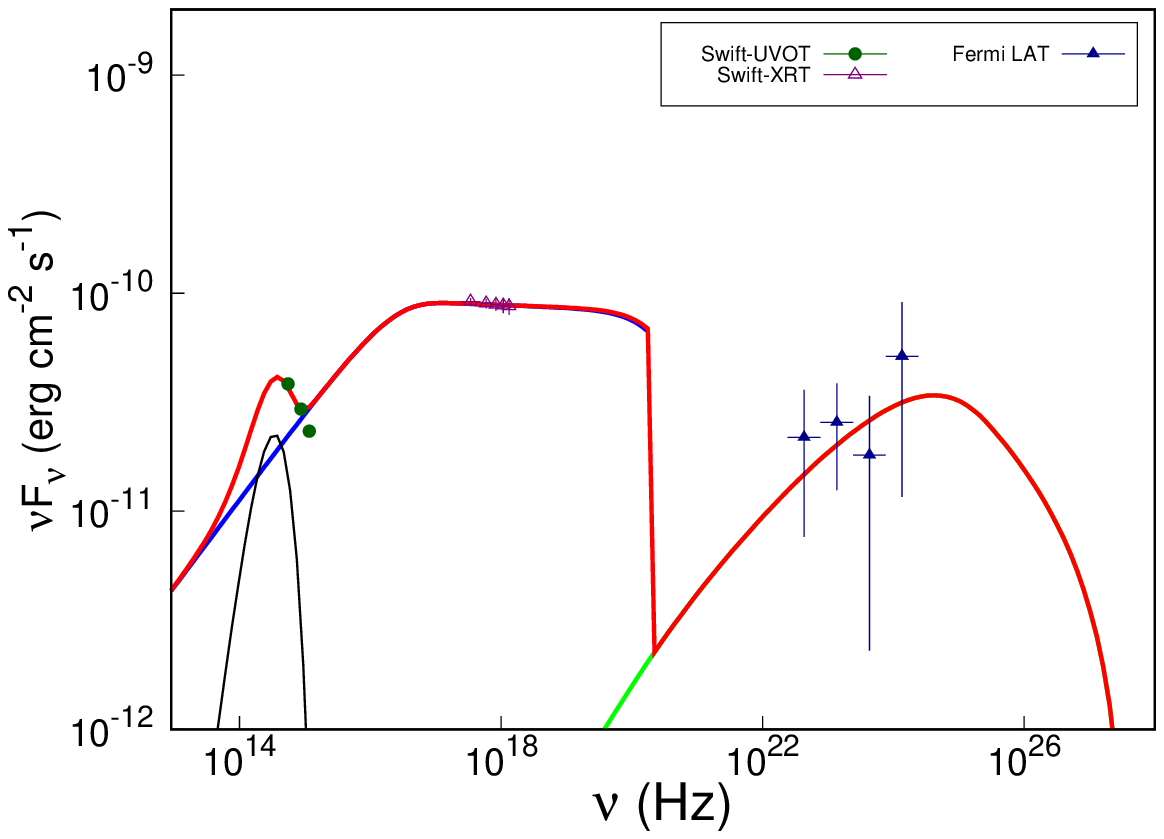}
      \includegraphics[scale=0.4,angle=270]{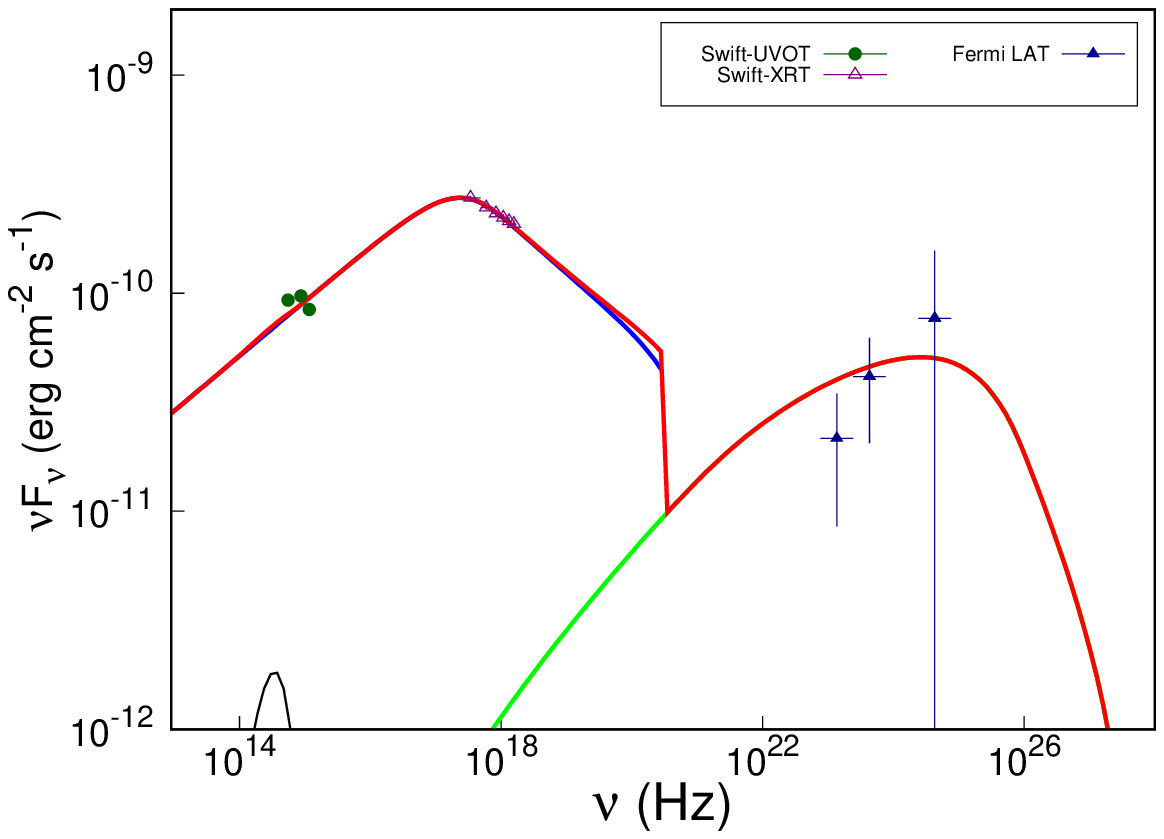}
     }
\caption{SEDs on 28 October 2022 and 14 August 2023 with added accretion disk emission component. The blue line refers to the synchrotron model, the green line refers to the SSC process, the black line refers to the accretion disk emission model and the red line refers to the sum of all the components.}
\label{figure-3}
\end{figure*}

However, during 14 August 2023, the source was in a high flux state and hence its optical/UV spectrum is dominated by the synchrotron emission from the jet.
The fit parameters are given in Table \ref{table-5} and the 
SEDs along with the model fits are shown in Figure \ref{figure-3}. 
 From a comparison of the physical parameters obtained from
the model fits to the SEDs of both the epochs, we noticed that the increased
X-ray polarization observed during the epoch of 14 August 2023 was associated with 
the increase in the bulk Lorentz factor of the relativistic jet of 1ES 1959+650. Nevertheless, the number of model parameters are larger than the information extracted from the SEDs and hence the parameters will be degenerate.

\begin{table}[htb]
\tabularfont
\caption{Results of the broadband SED analysis carried out for the epochs where the X-ray polarization is detected. Here, p and q are the particle indices while $\gamma_b$ is the break energy whereas $B$ is the magnetic field in Gauss. The emission region size (R) is given in log(cm).} \label{table-5} 
\begin{tabular}{lccr}
\topline
Parameter & 28 October 2022 & 14 August 2023   \\\midline
p & 2.2 & 2.5 \\
		q & 3.0 & 3.5 \\
		$\gamma_b$ &1.5$\times$10$^{4}$ & 3.4$\times$10$^{4}$ \\
    log (R) (cm) & 15 & 15 \\
		$\Gamma$ & 10& 28 \\
		B (G) & 2.1 & 2.2\\
\hline
\end{tabular}
\end{table}



\section{Discussion}
\label{sec:disc}
Observations with {\it IXPE} have enabled the detection of X-ray polarization
in the HSP blazar 1ES 1959+650. Out of the four epochs of observations, 
that span a duration of about 15 months, significant polarization was
detected on two epochs, namely 28 October 2022 and 14 August 2023, separated by about 10 months. On October 2022, in the 2$-$8 keV band, we measured the polarization of 
$\Pi_X$ = 9.0 $\pm$ 1.6\% and $\psi_X$ = 53 $\pm$ 5 deg, while on 14 August 2023,
we measured values of $\Pi_X$ = 12.5 $\pm$ 0.7\% and $\psi_X$ = 20 $\pm$ 2 deg. 
Thus, there is significant variation in the degree and angle of polarization between the two epochs. 
The high $\Pi_X$ measured on 14 August 2023, 
is similar to such high values measured for the blazars
1ES 0229+200 \citep{2023arXiv231001635E} with $\Pi_X$ = 17.9\% along $\psi_X$ $\sim$25 deg and 
the first {\it IXPE} observation of another HSP blazar Mrk 421 \citep{2022ApJ...938L...7D} with
$\Pi_X$ $\sim$15\%. From optical polarization observations carried out by the
Steward Observatory during the period 2008 and 2018, \cite{2023MNRAS.523.4504O} 
found the degree of optical polarization ($\Pi_o$) to range between 2.5\% $-$ 9\% with a 
mean value of $\Pi_o$ = 4.17 $\pm$ 0.12\% and a stable polarization angle.
The degree of X-ray polarization is thus higher than the average degree 
of polarization in the optical band. Such a trend is seen in 
other HSP blazars such as Mrk 501 \citep{2022Natur.611..677L,2024ApJ...970L..22H,2024arXiv240711128C}, Mrk 421 \citep{2022ApJ...938L...7D, 2023NatAs...7.1245D,2024AA...681A..12K}
and 1ES 0229+200 \citep{2023arXiv231001635E} based on simultaneous 
X-ray and optical polarization observations. Such energy dependent polarization with larger polarization degree in X-rays, when compared to that in optical suggests that
the X-ray emission must be produced by energetic electrons accelerated at shock 
fronts \citep{2020MNRAS.498..599T}. 

Recent X-ray spectral study of the source suggests the X-ray photon index to range from 1.34 $-$ 2.25 \citep{2023ApJ...951...94W}. 
Such a steep X-ray spectral index suggests that this energy range falls on the high 
energy end of the synchrotron component of the SED. Hence, this emission demands 
high electron energies as compared to the ones responsible for low energy 
emission at the optical waveband. If we assume these electrons are accelerated 
at the shock front under the Fermi process, then the high energy electrons 
populate in close vicinity to the shock front owing to their short cooling 
timescales. On the other hand, the longer cooling timescale of the low energy 
electrons lets them diffuse deep into the jet medium from the shock front. The 
shearing of the plasma at the shock front aligns the magnetic field \citep{1983MNRAS.202..553K,2022Natur.611..677L} 
which will result in strong polarised emission. Subsequently, the emission 
arising from the close vicinity to the shock front will be more polarized than 
the ones emerging from farther regions. The analysis of 1ES\,1959+650 suggests that the X-ray emission during both the epochs of high 
polarization are significantly larger than its optical polarization. This result 
supports that the efficient acceleration of electrons at the shock front is 
responsible for the broadband emission during these epochs. The two epochs of high X-ray polarization observed from 1ES\,1959+650 are 
separated by $\sim$10 months with significant variation in $\Pi_X$ and $\Psi_X$. 
Interestingly, the $\Pi_X$ increased by a factor of $\sim$1.4. We observed a variation in $\Psi_X$ from 53 $\pm$ 5 deg to 20 $\pm$ 2 deg.

The change in polarization angle could possibly be due 
to the changes in the position angle of the jet with time and such variation 
in the jet position angle is already known in 
blazars \citep{2018MNRAS.478.3199B,2020A&A...634A..87L,2022ApJS..260...12W}.
The high degree of polarization and modified polarization angle may also be associated with strong 
oblique shocks which are not normal to the jet flow, for example in collimation shocks \citep{2008Natur.452..966M}.
Also, we found that the observed broadband SED for the two epochs when X-ray polarization was detected can be well fit by the one zone leptonic emission model. The best fit model parameters obtained for both the epochs do not vary much except for the bulk Lorentz factor $\Gamma$. For the epoch of 14 August 2023, the observed $\Pi_X$ was about 1.4 times larger
than the value of $\Pi_X$ obtained on 28 October 2022. 
Incidentally, the bulk Lorentz factor of 14 August 2023 is about 2.8 times larger than the bulk Lorentz factor obtained on
28 October 2022. It is thus likely that the increase in $\Pi_X$ and decrease in
$\Psi_X$ on 14 August 2023 relative to 28 October 2022, could be 
associated to an increase in the bulk Lorentz factor of the jet 
and a change in the position angle of the jet in 1ES 1959+650, 
leading to changes in electric vector direction and hence $\Pi_X$.
The SED indicates that the flux was enhanced on August 14, 2023, compared to October 28, 2022. Our spectral modeling during this epoch suggests significant enhancement in the Doppler factor ($\delta$) of the jet. However, strong constraints cannot be imposed on $\delta$ since there exists a degeneracy between B and $\delta$ as shown by \cite{1998ApJ...509..608T}. Nevertheless, assumption of equipartition can effectively constrain B and then in turn $\delta$. The SED modeling does not suggest the presence of EC emission, where the target photons are boosted by $\Gamma$. Knowledge of $\delta$  and  $\Gamma$ could possibly be used to constrain the jet viewing angle.
If we attribute the flaring activity of the blazar as a result of a transient shock acceleration process, then the high Lorentz factor flow would produce strong shocks. Since, a shock tends to align the magnetic field lines \citep{1983MNRAS.202..553K}, high polarization detected during 14 August 2023 may indicate the presence of a strong shock. This is also consistent with the increased flux activity observed during this epoch.


\section{Conclusions}
\label{sec:conn}
We performed analysis of the X-ray polarimetric observations from {\it IXPE} on the source 1ES 1959+650. The observations were carried out on four epochs that span from May 2022 to August 2023.
The findings of the polarimetric study are as follows,
\begin{enumerate}
\item Of the four epochs of observations, we detected significant polarization on two
epochs, namely 28 October 2022 and 14 August 2023
\item From a model-independent analysis, we found a value of $\Pi_{X}$ = 9.0$\pm$1.6\% 
and $\Psi_{X}$ = 53 $\pm$ 5 deg in the 2$-$8 keV band for the 
observations on 28 October 2022. Similarly, from the observations on 14 August 2023,
using model independent analysis, we found $\Pi_X$ = 12.5 $\pm$ 0.7\% and 
$\psi_X$ = 20 $\pm$ 2 deg in the 2$-$8 keV band. We thus found polarization
to vary between the two epochs of observations.
\item From spectro-polarimetric analysis of model fits to the 2$-$8 keV spectra of 28 October 2022, 
we found values of $\Pi_{X}$ and $\Psi_{X}$ of 9.6$^{+0.03}_{-0.10}$ \% and 
49$^{+39}_{-32}$ deg, respectively. Similarly, for the observations done on 14 August 2023,
we found $\Pi_X$ = 12.6 $\pm$ 1.1\% and $\psi_X$ = 20 $\pm$ 2 deg. Thus, the 
polarimetric measurements obtained from
both model-independent and spectro-polarimetric analysis agree with each other, within the error limits.
\item To check for energy dependent polarization, we derived the polarization
parameters in different energy bands. For the epoch of 28 October 2022, 
polarization is significantly detected in the 2$-$3 keV, 3$-$5 keV and 2$-$8 keV bands. 
Both $\Pi_{X}$ and $\psi_X$ are found to be similar within error bars.
We thus did not find evidence of changes in polarization between energy bands. 
On the 14th of August 2023 observation, polarization is significantly detected 
in all energy bands, and while $\psi_X$ is found to be similar among energy bands, 
$\Pi_X$ is larger in the 5$-$8 keV band than that at 2$-$3 keV band, suggestive of the energy dependence in the polarization values.
\item Though optical polarization observations are not available during
the epoch of {\it IXPE} observations, the measured X-ray polarization during
the epochs of 28 October 2022 and 14 August 2023 is larger than the 
average optical polarization degree over a period of 10 years between 2008 and 2018. These observations tend to
favor a scenario of shock acceleration of electrons that 
produce the synchrotron optical and X-rays in 1ES 1959+650.
\item We found larger $\Pi_X$ on 14 August 2023 relative to 28 October 2022,
however, $\psi_X$ is lower on 14 August 2023 compared to 28 October 2022. This
could be associated with the variation in electric vector direction caused by the 
change in the position angle of the jet with time.
\item The broadband SEDs generated for the two epochs 28 October 2022 and 
14 August 2023 were well fit by a simple one zone leptonic emission model. 
The physical properties of the source derived for both the epochs are 
similar except for the bulk Lorentz factor, which, along with
the change in the jet position angle, could have caused the
change in the polarization behaviour of the source between the two epochs..
\end{enumerate}

With the results reported in this work, the number of blazars with 
X-ray polarization measurements from {\it IXPE} observations have 
increased to 10. X-ray polarimetric observations of more blazars with simultaneous observations at other wavelengths such as optical and radio are needed to better constrain the physical processes close to 
their central regions.





\section*{Acknowledgements}
We thank the anonymous referee for his/her critical comments.
The Imaging X-ray Polarimetry Explorer (IXPE) is a joint US and Italian mission.
 The US contribution is supported by the National Aeronautics and Space 
Administration (NASA) and led and managed by its Marshall Space Flight 
Center (MSFC), with industry partner Ball Aerospace (contract NNM15AA18C). The 
Italian contribution is supported by the Italian Space Agency (Agenzia Spaziale 
Italiana, ASI) through contract ASI-OHBI-2017-12-I.0, agreements 
ASI-INAF-2017-12-H0 and ASI-INFN-2017.13-H0, and its Space Science Data 
Center (SSDC) with agreements ASI- INAF-2022-14- HH.0 and ASI-INFN 2021-43-HH.0,
 and by the Istituto Nazionale di Astrofisica (INAF) and the Istituto Nazionale 
di Fisica Nucleare (INFN) in Italy. This research used data products provided by 
the {\it IXPE} Team (MSFC, SSDC, INAF, and INFN) and distributed with additional 
software tools by the High-Energy Astrophysics Science Archive Research Center 
(HEASARC), at NASA Goddard Space Flight Center (GSFC). 
Athira M Bharathan acknowledges the Department of Science and Technology (DST) for the INSPIRE Fellowship (IF200255). Special appreciation is extended to Soumya Gupta for her dedicated assistance.
\vspace{-1em}



\bibliography{JAA}{}
\end{document}